\documentclass[11pt]{article}
\usepackage{epsfig}
\usepackage{epstopdf}
\usepackage{amsmath}\usepackage{amssymb}
\usepackage{mathrsfs}
\usepackage{graphicx}

\newcommand {\be}{\begin{equation}}
\newcommand {\ee}{\end{equation}}
\newcommand {\ba}{\begin{eqnarray}}
\newcommand {\ea}{\end{eqnarray}}

\begin{document}

\def \a'{\alpha'}
\baselineskip 0.65 cm
\begin{flushright}
\ \today
\end{flushright}

\begin{center}{\large
{\bf Probing the top quark chromoelectric and chromomagnetic dipole moments in single top $tW$-channel at the LHC }} {\vskip 0.5 cm} {\bf Seyed Yaser Ayazi, Hoda Hesari and Mojtaba Mohammadi Najafabadi}{\vskip 0.5 cm
} School of Particles and Accelerators, Institute for Research in
Fundamental Sciences (IPM), P.O. Box 19395-5531, Tehran, Iran
\end{center}

\begin{abstract}
We study the effects of chromoelectric and chromomagnetic dipole moments (CEDM and CMDM) on the production cross section of single top $tW$-channel at the LHC based on the effective Lagrangian approach. We show that the impact of CEDM and CMDM could be large. Using the experimental measurement of the $tW$-channel cross section, constraints on CEDM and CMDM are extracted. These constraints are comparable with the ones obtained from the top pair analysis.

\end{abstract}

\section{Introduction}
The top quark is the heaviest Standard Model (SM) particle which has been discovered so far, and might
be the first place in which new physics effects could appear. While the top quark has been discovered  and studied in some details at the Tevatron, many of its properties are being studied at the LHC with high precision.

Since new physics can show up in the couplings of the top quark with other SM particles, in particular gauge bosons, the precise measurement of the top couplings among the top quark and gauge bosons is important. In particular, the top quark is copiously produced at the LHC, the LHC experiments are the places to probe these couplings\cite{Schilling:2012dx}.

At the LHC, top quarks are produced primarily via
two independent mechanisms. The dominant production mechanism is
the pair production processes $q\overline{q}\rightarrow
t\overline{t}$, $gg\rightarrow t\overline{t}$ and second is single
top production via electroweak interactions involving the $Wtb$ vertex. Single top quarks at the LHC are produced in three different modes: The s-channel (the involved $W$ is time-like), the t-channel mode (the involved $W$ is space-like), and the $tW$ production (the $W$ boson is real). Despite single top has a smaller cross section than top pair production, it can play an important role in top quark physics at LHC because this channel has potential to allow a direct measurement of $V_{tb}$ CKM matrix elements as well as its sensitivity to various new physics models. However sufficient integrated luminosity and improved method
of analysis can help us achieve detection of single top events
at the LHC. The first observable in single top study is the total cross section and measurement of any possible deviation from predicted value by the SM. Therefore, it is worthwhile to investigate the effects of physics beyond SM on single top quark
production.

As it is well-known SM has been successfully predicted experimental measurements with a great precision. Nevertheless, it is commonly accepted that it is a valid effective Lagrangian which is applicable at low energies. In many beyond SM theories which have been studied to date, reduction to the SM at low energies proceeds via decoupling of heavy particle with masses of order $\Lambda$. There have been many attempts to study the sensitivity of the LHC observables to various effective operators \cite{efL}. Our goal in this paper is to study the effect of anomalous couplings of the top quark with gluon via $tW$-channel of single top at the LHC. We will assume that new physics effects in $tW$ single top production are induced by consideration of an effective Lagrangian. Here, we confine our studies to interaction of mass dimension $5$ after spontaneous symmetry breaking. The total statistical and systematic uncertainties in the measurement of cross section for this process at the LHC is about $25\%$
with an integrated luminosity of $4.9~fb^{-1}$\cite{LHC tw}. The top quark is the heaviest quark, therefore effect of new physics on its coupling are expected to be larger than for any other fermions and deviation with respect to the SM predictions might be detectable.

The rest of this paper is organized as follows: In the next
section, Including the effective Lagrangian for $gt\overline{t}$ coupling, we calculate the analytical expression for the single $tW$ top cross section production at the LHC. In
section~3, we present the dependency of observables which we study at the LHC. Then, we find the allowed regions in parameters space of our effective Lagrangian and compare our results with the results obtained from observables in production of $t\bar{t}$ at LHC and EDM of top quark.  The conclusions are given in section~4.

\section{Framework and analytical calculations}

In this section, we introduce a model independent effective Lagrangian for the vertex of $gt\overline{t}$ and look for any possible deviation from SM prediction in production of $t\rm W $-channel at the LHC. In this approach, we assume that SM modified by an addition 5-dimensional Lagrangian which include interaction of top pair and  gluon and consider the effect of this Lagrangian in production of $tW$ single top production. Coefficients of this Lagrangian parameterize the low energy effects of the underlying high scale physics. As it is well-known, the gauge invariant effective Lagrangian for the interactions between the top quark and gluons which include the CEDM and CMDM form factors is given by:
\begin{eqnarray}\label{lag}
\mathcal{L}_{eff}=g_{s}\bar{t}\lambda^{a}\gamma^{\mu}t G^{a}_{\mu}+g_{s}\bar{t}\lambda^{a}\frac{\sigma^{\mu\nu}q_{\mu}}{4m_{t}}[F_2(q^2)+i\gamma^5F_3(q^2)]t G^{a}_{\nu}
\end{eqnarray}
where $\lambda^{a}(a=1,\ldots,8)$ are the $SU(3)_{C}$ color matrices and $ F_{2}\left( q^{2}\right)$ and $F_{3}\left( q^{2}\right)$ are, respectively, the CMDM and CEDM forms factors of the top quark. Notice that a sizable non-zero CEDM would be the signal of a new type of CP-violating interaction beyond CKM phase and can contribute to electric dipole moment (EDM) of neutron. For this reason, experimental upper bound on EDM of neutron constraint this coupling \cite{CEDM},\cite{TOP EDM}. In the following, we use this constraint on parameters space of above Lagrangian and compare these bounds with our result which arise from $t\rm W $ single top production at LHC.

Assuming $ \vert q^{2} \vert\ll\Lambda$, where $\Lambda$ is the scale of new physics, the form factors can be approximated by
\begin{eqnarray}\label{2}
F_{3}( q^{2})\approx\widetilde{\kappa} \qquad  F_{2}\left(q^{2}\right)\approx \kappa \qquad    for \quad \vert q^{2} \vert\ll\Lambda
\end{eqnarray}
where $\widetilde{\kappa}$ and $\kappa$ are independent of $q^{2}$ . The CMDM of the top quark is then given by $\frac{g_{s}\kappa}{( 2m_{t})}$,while CEDM is $\frac{g_{s}\widetilde{\kappa}}{\left( 2m_{t}\right) }$.

\begin{figure}[t]
\epsfysize=1.6in
\centerline{\epsfbox{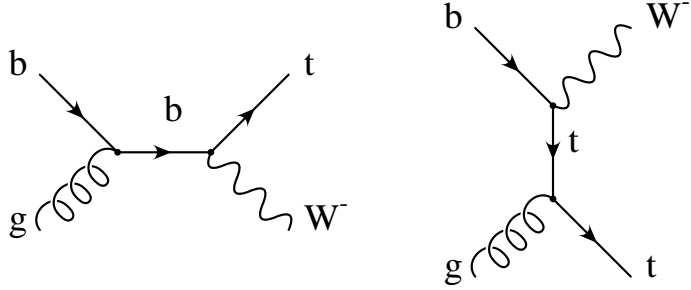}}
\caption{Feynman diagrams for the $t \rm W$ channel of single top}
\label{twfig}
\end{figure}

As it is shown in Fig.~\ref{twfig}, $gt\overline{t}$ effective Lagrangian can contribute to the right diagram in $tW$ of single top production at the LHC. We calculate the amplitude of $bg\rightarrow t\rm W$ process including CEDM and CMDM effect.  This amplitude is given by:
\begin{eqnarray}\label{matrixelaement}
\overline{|M^2|} &=& \frac{g^2g_s^2}{192}{\vert V_{tb} \vert}^{2}[\frac{f_1}{(m_t^2-\hat{u})^2}+\frac{m_{\rm W}^2f_2}{\hat{s} (m_t^2-\hat{u})^2 (m_t^3-\hat{u} m_t)^2} \\ \nonumber
&+&\frac{f_3}{\hat{s} m_t^2 (m_t^2-\hat{u}){}^2 m_{\rm W}^2}+\frac{32m_{\rm W}^4}{\hat{s}(m_t^2-\hat{u})}+\frac{\tilde{\kappa }^2f_4}{m_t^2 m_{\rm W}^2}+\frac{16 \hat{s} \hat{u}f_5}{(m_t^2- \hat{u})^2}\\ \nonumber
&+&\frac{\tilde{\kappa }f_6}{\hat{s}(m_t^2- \hat{u}) m_{\rm W}^2} ]
\\ \nonumber
\end{eqnarray}
where $g$ and $g_{s}$ are respectively weak and strong coupling constants and $m_t$, $ m_{\rm W}$  are the masses of top and W gauge boson and $V_{tb}$ is  CKM matrix element. The explicit forms for $f_i$ ($i=1,..6$)  in terms of Mandelstam variables, $\kappa$ and $\tilde{\kappa}$  are given in
Eqs.~(\ref{f1}-\ref{f6}) of the appendix.

\section{Sensitivity analysis for anomalous $gt\overline{t}$ couplings}

In this section, we study the total cross section of $tW$ single top production at the LHC and
study the effect of CEDM and CMDM couplings on it.

Recently, $\rm CMS$ collaboration reported  the measured value of cross section of $tW$ single top production at the center-of-mass energy  $\sqrt{S}=7 ~\rm TeV$  with an integrated luminosity of $4.9~\rm fb^{-1}$\cite{LHC tw}:
\begin{eqnarray}
\sigma_{\rm LHC}(pp\rightarrow t{\rm W} ) & =
16^{+5}_{-4}~[pb].\label{exp}
\end{eqnarray}

This measurement is in agreement with the SM expectation $15.6\pm0.4^{+1}_{-1.2}$ \cite{SM prediction}.
The hadronic cross section for production of $tW$ can be obtained by integrating over the parton level cross section convoluted with the parton distribution functions:
\begin{eqnarray}
\sigma(pp\rightarrow t\rm W) & = &\sum_{ab} \int
dx_1dx_2f_a(x_1,Q^2)f_b(x_2,Q^2) \widehat{\sigma}(ab\rightarrow
t\rm W), \
\end{eqnarray}
where $f_{a,b}(x_i,Q^2)$ are the parton structure functions of
proton. The parameters  $x_1$ and $x_2$ are the parton momentum fractions and $Q$
is the factorization scale.

\begin{figure}
\begin{center}
\centerline{\vspace{-1cm}}
\centerline{\includegraphics[scale=0.3]{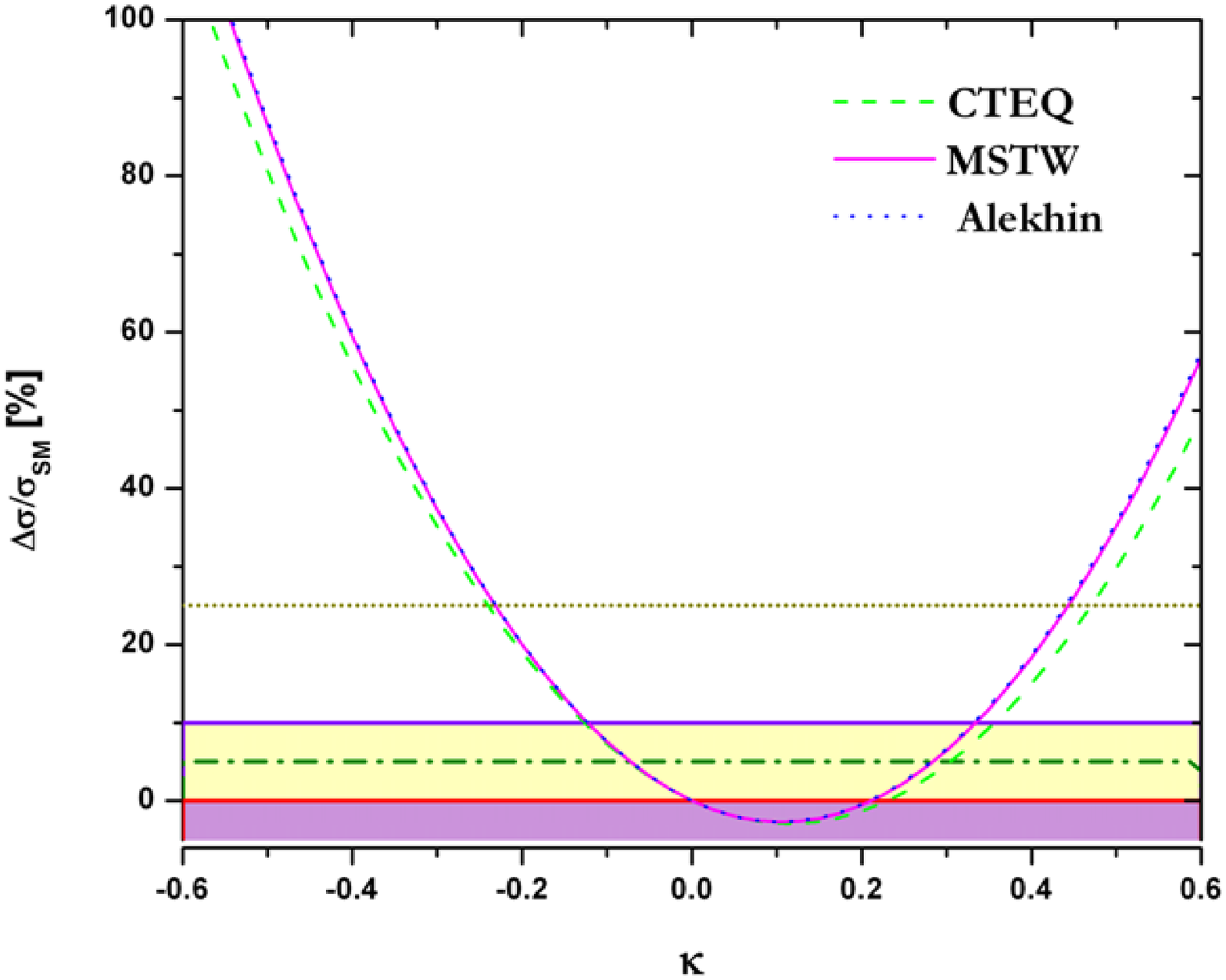}\hspace{-3mm}\includegraphics[scale=0.3]{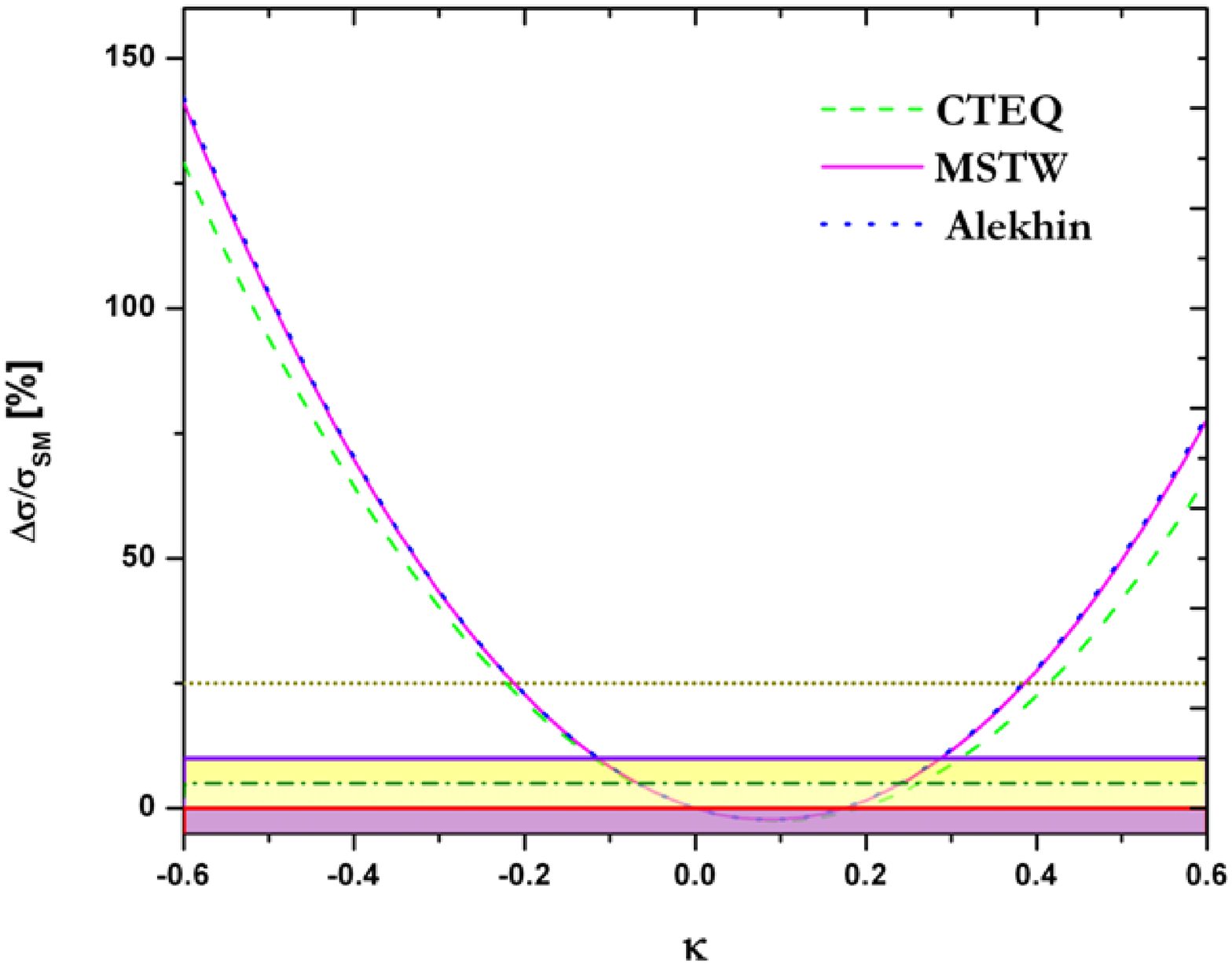}}
\centerline{\vspace{1cm}\hspace{0.5cm}(a)\hspace{7cm}(b)}
\centerline{\vspace{-2.5cm}}
\end{center}
\caption{a) $\Delta\sigma(pp\rightarrow
t\rm W)$/$\sigma_{SM}$ versus $\kappa$. In this figure, we have set
$\sqrt{S}=7~ \rm TeV$. The green curve (dashed), the pink curve (line) and blue curve (dotted) respectively
correspond to CTEQ6.6M \cite{CTEQ}, MSTW2008 \cite{MSTW} and
ALEKHIN2 \cite{Alekhin} structure functions. b) Similar to Fig.~a
except that $\sqrt{S}=10~ \rm TeV$. The horizontal small dotted (yellow dark), violet and dot-dashed (green) lines respectively
 correspond to $\pm 25\%$, $\pm 10 \%$ and $\pm 5\%$ uncertainties in
the measurement of $\sigma(pp\rightarrow
t\rm W)$.}\label{deltacrossk}
\end{figure}

\begin{figure}
\begin{center}
\centerline{\vspace{-1.2cm}}
\centerline{\includegraphics[scale=0.3]{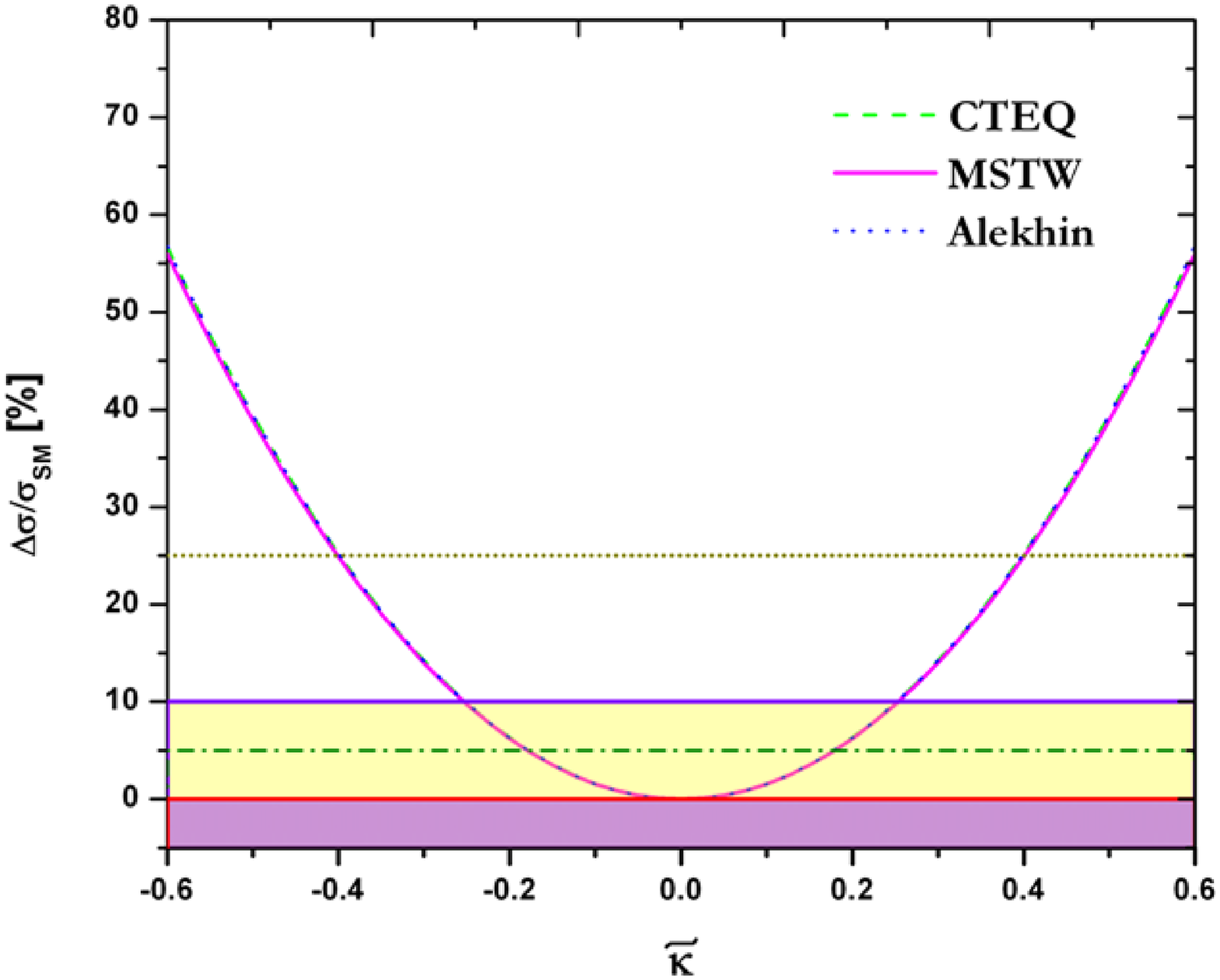}\hspace{-3mm}\includegraphics[scale=0.3]{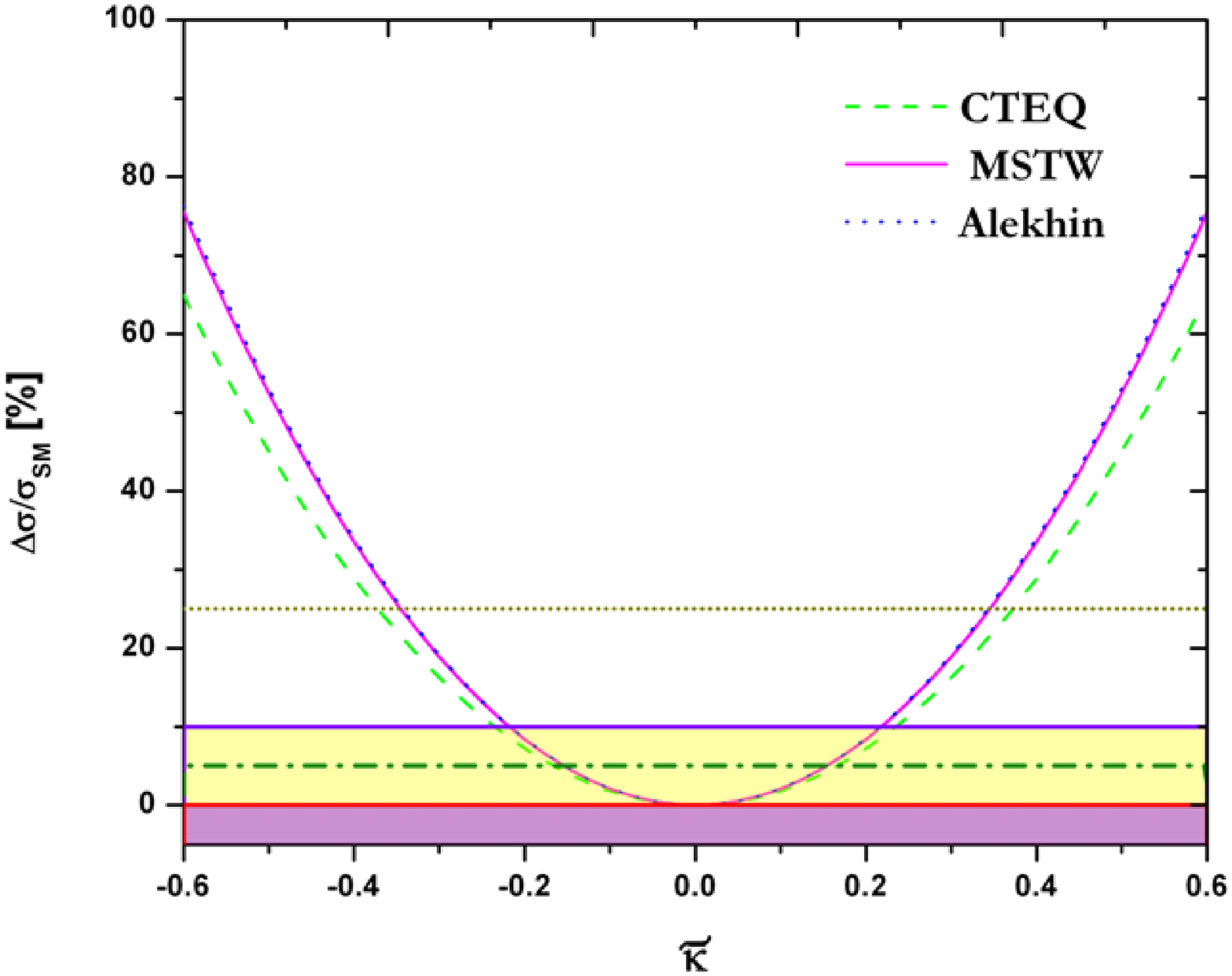}}
\centerline{\vspace{1cm}\hspace{0.5cm}(a)\hspace{7cm}(b)}
\centerline{\vspace{-2.5cm}}
\end{center}
\caption{ a) $\Delta\sigma(pp\rightarrow
t\rm W)$/$\sigma_{SM}$ versus $\widetilde{\kappa}$ for the center-of-mass energy of $7~\rm TeV$ (left) and $10~\rm TeV$ (right). Input parameters are similar to Fig.~\ref{deltacrossk}.}\label{deltacrosskt}
\end{figure}

In this paper, we obtain the direct constraints on dipole operators including top quark in the above effective Lagrangian approach.
We consider the total cross section of $tW$ single top production at the LHC and
study the effect of CMDM and CEDM coupling on it. For this study, we consider the relative change in cross section
which is defined as:
\begin{eqnarray}
 R=\frac{\Delta\sigma}{\sigma_{SM}}=\frac{\sigma-\sigma_{SM}}{\sigma_{SM}}, \
\end{eqnarray}
where $\sigma$ is total cross section in the presence of CMDM and CEDM couplings.
The relative change in cross section of the single top production at
the LHC are shown in Fig.~\ref{deltacrossk} and \ref{deltacrosskt}.

In these figures, we consider that only CMDM or CEDM coupling exists and display  the relative change in cross section of
$\sigma(pp\rightarrow t\rm W)$ versus $\kappa$ and $\widetilde{\kappa}$. In Fig.~\ref{deltacrossk}-a, we have set $\sqrt{S}=7~ \rm TeV$.  In Fig.~\ref{deltacrossk}-b, as explained in the caption, we
have set $\sqrt{S}=10~ \rm TeV$ which might be measured in ongoing run of the LHC. In the center of mass of energy
$10~ \rm TeV$, leading order SM cross section of $\sigma(pp\rightarrow
t\rm W)$ have been obtained $38.8~\rm pb$ \cite{SM prediction}. To calculate
$\sigma(pp\rightarrow t\rm W)$, we have used the CTEQ6.6M
\cite{CTEQ}, MSTW2008 \cite{MSTW} and ALEKHIN2
\cite{Alekhin} as the parton structure functions (PDF). The green
curve (dashed), the pink curve (line) and blue curve (dotted)
are corresponding to CTEQ6.6M, MSTW2008 and ALEKHIN2
structure functions, respectively.

An interesting observation from Fig.~\ref{deltacrossk} and \ref{deltacrosskt} is that the correction to $tW$-channel cross section due to CMDM and CEDM is  sensitive to the choice of parton distribution function, in particular at large values of  $\kappa$ and $\widetilde{\kappa}$.
As it is seen in these figures, different
structure functions change the value of
$\sigma(pp\rightarrow t\rm W)$ more than $10\%$ for large values of $\kappa$ and $\widetilde{\kappa}$. Considering the CTEQ PDF, the presence of  $\kappa$ or $\widetilde{\kappa}$ can change total cross section more than $10~\rm pb$. At small value in the range of $[-0.2,0.2]$, $\Delta\sigma(pp\rightarrow t\rm W)$/$\sigma_{SM}$ is almost robust against the choice of PDF.

 Fig.~\ref{deltacrossk} demonstrates that the effect of the presence of
CMDM can change the total cross section more than $40 \%$ for  $ \kappa=0.4$. The horizontal small dotted (yellow dark), violet and dot-dashed (green) lines respectively correspond to $\pm 25\%$, $\pm 10 \%$ and $\pm 5\%$ uncertainty in
the measurement of $\sigma(pp\rightarrow
t\rm W)$. As it was mentioned, the total statistical and systematic uncertainties in the measurement of cross section for $tW$ single top production at the LHC is about $25\%$
with an integrated luminosity of $4.9~fb^{-1}$ at center-of-mass energy $7~ \rm TeV$. With $10\%$ uncertainty  in future measurement of the total cross
section,  from Fig.~\ref{deltacrossk} (\ref{deltacrosskt}), we can put upper bound on $\kappa$ ($\widetilde{\kappa}$) down to $0.14$ ($0.23$). If in forthcoming run of LHC (in $10~\rm TeV$), we measure the total cross section  even with $5\%$ uncertainty, we can constrain $\kappa$ ($\widetilde{\kappa}$) down to $0.09$ ($0.15$). As it is seen in Fig.~\ref{deltacrosskt}, in spite of CMDM ($\kappa$), the cross section is symmetric with respect to $\widetilde{\kappa}$ because  CEDM coupling enters in the cross section in even powers when $\kappa=0$. Therefore, in this situation, the cross section is not a CP violating observable.

Another observation from Fig.~\ref{deltacrossk} is that the effect of presence of
CMDM ($\kappa$), in spite of CEDM ($\widetilde{\kappa}$),  can be destructive. If the total uncertainties in the measurement of cross section  is less than $5\%$, we can distinguish between CEDM and CMDM effects.

\begin{figure}
\begin{center}
\centerline{\hspace{0.5cm}\epsfig{figure=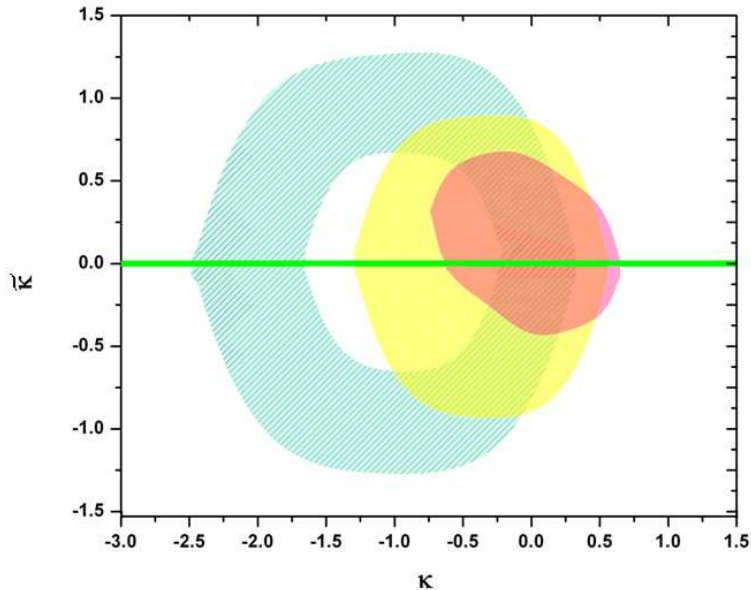,width=12cm}}
\centerline{\vspace{-2.1cm}}
\end{center}
\caption{ Red area depicts ranges of parameters space in CMDM ($\kappa$) and  CEDM ($\widetilde{\kappa}$) couplings plane for which prediction of gluonic dipole effective Lagrangian on $tW$ single top production at LHC are consistent with experimental measurements.
 Green line shows neutron EDM constraint on $\widetilde{\kappa}$. Hatched cyan (Yellow) shaded area depicts allowed region which are consistent with $t\overline{t}$ total cross section(spectrum measurement of $m_{t\overline{t}}$) at LHC. Shaded area which depict $t\overline{t}$-channel constraints have been borrowed from \cite{TOP EDM}.} \label{scater}
\end{figure}

In Figs.~\ref{scater}, red area depicts ranges of parameters space in CMDM ($\kappa$) and  CEDM ($\widetilde{\kappa}$) couplings plane for which prediction of effective Lagrangian (equation \ref{lag}) on $tW$ single top production at LHC are consistent with experimental measurements.
When performing such study, one should take into account constraints from other studies. There exist many direct and indirect constraints on dipole operator.
Presently, the most sensitive observable obtained from mercury  and neutron EDM. In \cite{TOP EDM}, it is shown that the neutron EDM constrains the top CEDM to be $\widetilde{\kappa}<8\times 10^{-3}$. Moreover, the constraints from $d_{Hg}$ and  the top electric dipole moment, provide weaker bound on $\widetilde{\kappa}$. Furthermore, the CEDM and CMDM couplings of the top quark directly affect on top pair production at hadron colliders \cite{tt constriants}.
We have borrowed the results of LHC constraints which come from $t\overline{t}$ total cross section on CEDM and CMDM of top from reference \cite{TOP EDM}. These results have been shown in Fig.~\ref{scater}. In this figure green line depicts neutron EDM constraint on $\widetilde{\kappa}$. Hatched cyan shaded area depicts the allowed region which is consistent with $t\overline{t}$ total cross section at LHC. Yellow area shows allowed region which is consistent with spectrum measurement of $m_{t\overline{t}}$ at LHC. It is remarkable that the allowed regions of top pair production cross section and spectrum measurement of $m_{t\overline{t}}$ at the LHC  overlap with allowed region of $tW$ single top production at LHC. As it can be seen in this figure, $tW$ single top constraints on $\kappa$ and $\widetilde{\kappa}$ are stronger than direct constraints which come from  $t\overline{t}$ total cross section and spectrum measurement of $m_{t\overline{t}}$ at LHC.

\section{Concluding remarks}
In this paper, based on the effective Lagrangian approach, we modify the SM  by the additional 5-dimensional operators which include the interaction of top quark with  gluon and consider the effect of this Lagrangian in production of $tW$ single top. Coefficients of this Lagrangian are related to CEDM and CMDM form factors. We consider the total cross section of $tW$ single top production at the LHC and study the effect of CEDM and CMDM couplings on it. We have found the allowed regions in parameters space of CEDM and CMDM in such a way that the experimental measurement of $\sigma(pp\rightarrow t\rm W)$ is   satisfied. We also investigate the effect of different PDFs on $tW$ single top production at LHC as a function of $\kappa$ and $\widetilde{\kappa}$.

We have shown that deviation of the $tW$-channel single top cross section from the SM value is significant. We consider constraints on CEDM and CMDM couplings which come from $t\overline{t}$ total cross section and spectrum measurement of $m_{t\overline{t}}$  at the LHC and compare them with our results.
We have shown that constraints on $\kappa$ and $\widetilde{\kappa}$ which arise from $tW$ single top production at LHC are comparable with the ones coming from $t\overline{t}$ production at the LHC. It is notable that with the current $tW$ cross section precision measurement, tight bounds are obtained and therefore with more precision measurements in future even more stringent limits than $t\overline{t}$ cross section could be achieved.

\section{Acknowledgement}
We would like to thank H. Kanpour for helping us in technical issues.
\section*{Appendix}
Here, we list the formulas of $f_i$ which have been applied in calculation of single top production cross section.
Notice that mass dimension of $f_i$ are not equal.
\begin{eqnarray}\label{f1}
\nonumber
f_1&=&16 \hat{s} m_t^2-16 \kappa  \hat{s} m_t^2+20 \kappa ^2 \hat{s} m_t^2-16 \hat{u} m_t^2+24 \kappa ^2 \hat{u} m_t^2\\
&-&10 \kappa ^2 m_t^4-16\hat{u} m_t^4+24 \kappa  m_t^4+32 \hat{u}^2 m_t^2
\end{eqnarray}
\begin{eqnarray}\label{f2}
\nonumber
f_2&=&(-4 \hat{s} \kappa ^2-32 \hat{s} \kappa +16 \hat{u}) m_t^8+\hat{u} ((16 \kappa ^2+96 \kappa +32) \hat{s}+48 \hat{u}) m_t^6\\ \nonumber&+&\hat{u}^2 \left(\left(-24 \kappa ^2-96 \kappa -64\right) \hat{s}-80\hat{u}\right)m_t^4 +\hat{u}^3 \left(\left(16 \kappa ^2+32 \kappa +32\right) \hat{s}+32 \hat{u}\right) m_t^2\\ &-&4 \kappa ^2 \hat{s} \hat{u}^4-16 m_t^{10}\\ \nonumber
\end{eqnarray}
\begin{eqnarray}\label{f3}
\nonumber
f_3&=&8 m_t^{10}+\left(-16 \hat{s}-8 \hat{u}\right) m_t^8+\left(8 \hat{s}^2+\left(10 \kappa ^2-16 \kappa +16\right) \hat{u} \hat{s}+8 \hat{u}^2\right) m_t^6\\ \nonumber
&+&\hat{u} \left(\left(-10 \kappa ^2+16 \kappa -8\right) \hat{s}^2-8 \hat{u}^2+\left(-20 \kappa ^2+16 \kappa -16\right) \hat{s} \hat{u}\right) m_t^4\\
&+&\ \left(12 \hat{s}+10\hat{u}\right)\kappa  \hat{s} \kappa  \hat{u}^2 m_t^2-2 \kappa ^2 \hat{s}^2 \hat{u}^3\\ \nonumber
\end{eqnarray}
\begin{eqnarray}\label{f4}
f_4&=&-4 m_w^4+\left(4 \hat{s}+4 \hat{u}\right) m_w^2-2 \hat{s} \hat{u}+m_t^2 \left(2 \hat{u}-2 m_{\rm W}^2\right)\\ \nonumber
\end{eqnarray}
\begin{eqnarray}\label{f5}
\nonumber
f_5&=&-16 \hat{s} \hat{u}-16 \kappa  \hat{s} \hat{u}-24 \kappa ^2 \hat{s} \hat{u}-24 \kappa  \hat{u}^2-18 \kappa ^2 \hat{u}^2\\
&-&\frac{16 \hat{u}^3}{\hat{s} }+\frac{4 \kappa ^2 \hat{s} \hat{ u}^2}{m_t^2}+\frac{4 \kappa ^2 \hat{u}^3}{m_t^2}\\ \nonumber
\end{eqnarray}
\begin{eqnarray}\label{f6}
f_6&=&\kappa  \hat{s} [\left(8\hat{u}-8 m_{\rm W}^2\right) m_t^2+\hat{u} \left(-8 \hat{s}-8 \hat{u}\right)+\left(16 \hat{s}+8 \hat{u}\right) m_{\rm W}^2]
\end{eqnarray}
where $s,t$ and $u$ are Mandelstam variables, $m_t$, $ m_{\rm W}$  are  the masses of top and W gauge boson. The parameters $\widetilde{\kappa}$ and $\kappa$ are, respectively CEDM and CMDM couplings.

\end{document}